\documentclass[reprint,...]{revtex4-1}
\usepackage[utf8]{inputenc}
\usepackage{amsmath,amssymb,epsfig}
\usepackage{graphicx}

\usepackage{multirow}
\usepackage{url}
\usepackage{color}

\usepackage{float}
\usepackage{wrapfig}
\usepackage[hyperfootnotes=true]{hyperref}

\usepackage{lipsum}

\usepackage{bm}        
\usepackage{dcolumn}   

\def\beq{\begin{equation}}
\def\eeq{\end{equation}}
\def\bea{\begin{eqnarray}}
\def\eea{\end{eqnarray}}

\DeclareMathOperator{\ev}{eV}
\DeclareMathOperator{\gev}{GeV}

\DeclareMathOperator{\dec}{dec}
\DeclareMathOperator{\br}{BR}

\newcommand{\be}{\begin{equation}}
\newcommand{\ee}{\end{equation}}

\newcommand{\ba}{\begin{array}}
\newcommand{\ea}{\end{array}}

\newcommand{\bal}{\begin{align}}
\newcommand{\eal}{\end{align}}
\newcommand{\bi}{\begin{itemize}}
\newcommand{\ei}{\end{itemize}}
\newcommand{\ben}{\begin{enumerate}}
\newcommand{\een}{\end{enumerate}}
\newcommand{\bc}{\begin{center}}
\newcommand{\ec}{\end{center}}
\newcommand{\bt}{\begin{table}}
\newcommand{\et}{\end{table}}
\newcommand{\btb}{\begin{tabular}}
\newcommand{\etb}{\end{tabular}}
\newcommand{\bs}{ \begin{split}}
\newcommand{\es}{\end {split}}

 \newcommand{\bes}{} 
  \def\bes#1\ee{\begin{equation}#1\end{equation}}

\newcommand{\nn}{\nonumber}

\newcommand{\RN}[1]{%
  \textup{\uppercase\expandafter{\romannumeral#1}}%
}

\begin{document}

\thispagestyle{empty}

\title{Light Dark Matter from Leptogenesis}
\author{Adam Falkowski}\email{adam.falkowski@th.u-psud.fr}
\affiliation{Laboratoire de Physique Th\'{e}orique, CNRS, Univ. Paris-Sud, Universit\'{e}  Paris-Saclay, 91405 Orsay, France}
\author{Eric Kuflik}\email{eric.kuflik@mail.huji.ac.il}
\affiliation{Racah Institute of Physics, Hebrew University, Jerusalem 91904, Israel}
\author{Noam Levi}\email{noam@mail.tau.ac.il, noam@ias.edu}
\author{Tomer Volansky}\email{tomerv@post.tau.ac.il}
\affiliation{Raymond and Beverly Sackler School of Physics and Astronomy, Tel-Aviv University, Tel-Aviv 69978, Israel}
\affiliation{School of Natural Sciences, The Institute for Advanced Study, Princeton, NJ 08540, USA}

\date{\today}
\begin{abstract}
We consider the implications  of a shared production mechanism between the baryon asymmetry of the universe and the relic abundance of dark matter, that does not result in matching asymmetries. We present a simple model within a two sector leptogenesis framework, in which right handed sterile neutrinos decay out of  equilibrium to both the Standard Model and the dark sector, generating an asymmetry in one and populating the other. This realization naturally accommodates light dark matter in the keV mass scale and above. Interactions in the dark sector may or may not cause the sector to thermalize, leading to interesting phenomenological implications, including hot, warm or cold thermal relic dark matter, while evading cosmological constraints. Under minimal assumptions the model provides a novel non-thermal production mechanism for sterile neutrino dark matter and predicts indirect detection signatures which may address the unexplained 3.5 keV line observed  in various galaxy clusters.
\end{abstract}

\maketitle

\section{Introduction}\label{sec1}

The existence of Dark Matter (DM), the baryon asymmetry of the universe and neutrino masses are major observational evidence for physics beyond the Standard Model (SM).
Attempts at explaining these observations over the past decade, both separately and simultaneously, have given rise to various DM models, 
with DM typically residing at or above the GeV mass scale.
One such   class of models  
which relates the baryon asymmetry with dark matter  is known as Asymmetric Dark Matter (ADM)
~\cite{Nussinov:1985xr, Kaplan:1991ah,Farrar:2004qy,Hooper:2004dc,Kitano:2004sv,Agashe:2004bm,Kitano:2008tk,Nardi:2008ix,Kaplan:2009ag,Cohen:2010kn,Hall:2010jx,Feldstein:2010xe,Shelton:2010ta, Davoudiasl:2010am,Haba:2010bm,Gu:2010ft,Blennow:2010qp,McDonald:2011zza,Cheung:2010gj,Cheung:2010gk,Dutta:2010va,Zurek:2013wia}. 
Motivated by the observation that $\Omega_{\rm DM} \simeq 5~\Omega_{B} $, these models generically provide a single production mechanism that generates 
comparable number densities of baryons and DM in the early universe, leading to the prediction $m_{\rm DM} \sim 5~m_{\rm proton}$.    
More generally a broader range of masses can be achieved within the ADM framework
~\cite{Nussinov:1985xr,
Cohen:2009fz,
Falkowski:2011xh,
Buckley:2010ui}. 

A crucial ingredient of ADM is that the symmetric dark matter component can be
efficiently annihilated away, rendering the overall density asymmetric. 
Then a  natural question to ask is: 
{\it what are the predictions for dark matter in the case where its annihilation cross-section is  
not large enough to remove the symmetric component?} 
Such a scenario corresponds to a significant fraction of the models' parameter space and predicts symmetric and asymmetric densities in the dark and visible sectors, respectively. 
An interesting framework that allows one to address the above question is leptogenesis
~\cite{Fukugita:1986hr}, (for a review see, e.g.~\cite{Davidson:2008bu}).  In leptogenesis, a lepton asymmetry is first generated in the early universe by out of equilibrium, CP-violating decays of right-handed neutrinos.  The asymmetry is then converted to a baryon asymmetry using $B$+$L$-violating interactions, active before the EWSB phase transition.
A realization that connects ADM with this framework is the two-sector leptogenesis framework~\cite{Falkowski:2011xh},
 in which the sterile neutrinos couple (in addition to SM leptons in the visible sector) to a dark sector, thereby generating an asymmetry in both the visible and dark sectors simultaneously.

In this work, we take a first step in answering the above question by studying the two sector leptogenesis scenario  
without matching asymmetries between the two sectors (for other possible scenarios, see~\cite{McDonald:2011sv,DEramo:2011dhr,McDonald:2011zza,Cui:2011ab}).
 Since the production of an asymmetry requires CP violation and is therefore loop suppressed, relaxing the asymmetric requirement in the dark sector generically results in  a tree level enhancement in production, leading to $n_{\rm DM}\gg n_{B}$ and therefore $m_{\rm DM} \ll m_{\rm proton}$.   Thus, this scenario predicts light dark matter.   Theories of light dark matter, in the keV to GeV mass range, have received significant attention in recent years~\cite{
Boehm:2003hm,
Pospelov:2007mp,
Hooper:2008im,
Feng:2008ya,
Morrissey:2009ur,
Hall:2009bx,
Essig:2010ye,
Kang:2010mh,
Nelson:2011sf,
Carena:2011jy,
Chu:2011be,
Lin:2011gj,
An:2012va,
Hooper:2012cw,
Ho:2012br,
Tulin:2012wi,
Essig:2013lka,
Tulin:2013teo,
Zurek:2013wia,
Hochberg:2014dra,
Heeck:2014zfa,
Lindner:2010wr,
Boddy:2014yra,
Choi:2015bya,
Lee:2015gsa,
Altmannshofer:2014vra,
Hochberg:2014kqa,
DAgnolo:2015ujb,
Izaguirre:2015yja,
Kuflik:2015isi,
Graham:2015rva,
DiFranzo:2015nli,Bernal:2015ova,
Bernal:2015xba,
Farina:2016llk,
Pappadopulo:2016pkp,
Kopp:2016yji,
Boddy:2016bbu, 
Chacko:2016kgg,
Dror:2016rxc,
Liu:2016qwd,
Agrawal:2016quu,
Buen-Abad:2017gxg,
Battaglieri:2017aum,
Heeck:2017xbu,
Berlin:2017ftj,
DAgnolo:2017dbv,
Tulin:2017ara,
Green:2017ybv,
Zhao:2017wmo,
Allahverdi:2017edd}.

Allowing for interactions in the dark sector leads to several interesting implications, including a thermal hidden sector.   Moreover, under mild assumptions the DM plays the role of a sterile neutrino and the scenario provides a new production mechanism for such a  dark matter candidate (for a review and references, see~\cite{Kusenko:2009up, Adhikari:2016bei}).  Correspondingly, indirect detection signatures are predicted and allows  for an explanation of the 3.5 keV line seen in various X-ray observations~\cite{Bulbul:2014sua,Boyarsky:2014jta,Abazajian:2017tcc}. 

\begin{figure}
\includegraphics[width=60mm]{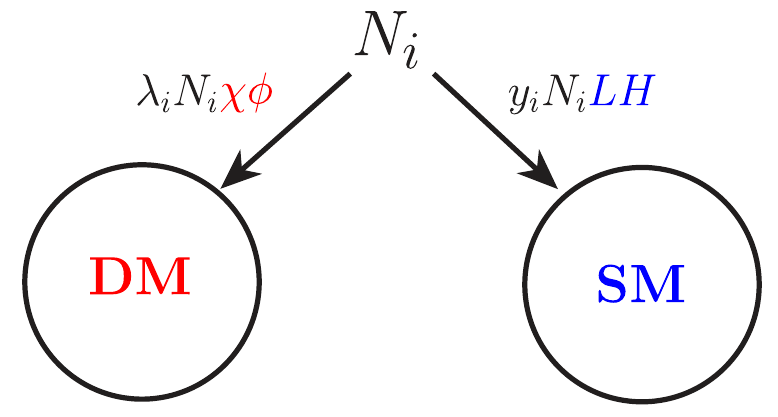}
\caption{\label{fig:scheme} A schematic view of the two sector leptogenesis framework.  The SM and DM sectors are indirectly connected via Yukawa interactions with the same heavy right-handed neutrinos, $N_i$. The complex couplings, $y_i$, lead to CP violation in $N_i$ decays, and consequently particle-antiparticle asymmetry for the SM leptons.  The couplings to the dark sector, $\lambda_i$, may or may not be complex.}
\end{figure}

Our framework consists of  two separate sectors. A Standard Model (SM) sector and a hidden (``dark'') sector, coupled via Yukawa interactions with 3 generations of heavy right-handed sterile neutrinos, $N_{i}$ $(i=1,2,3)$, as illustrated in Fig.~\ref{fig:scheme}. We consider the case where Yukawa interactions between the sterile neutrinos and the SM dynamically trigger leptogenesis at early epochs of the universe, and the sterile neutrino decays to the DM sector produce a relic abundance of symmetric dark matter.
To be concrete, consider a minimal two sector leptogenesis model~\cite{Falkowski:2011xh}
\beq 
-\mathcal{L}_{\rm int}  \supset \frac{1}{2}M_N N^2 + m_\chi \bar{\chi}\chi + y h L N  +\lambda \phi N {\chi}  + \textrm{h.c.} \label{lag}\,.
\eeq
Here $N$ represents the right-handed neutrino, $L$ is the SM lepton doublet, $H$ is the Higgs doublet,  $\phi$ is a scalar field residing in the dark sector, and $\chi$ is the fermionic DM candidate with a Dirac mass term.  
We neglect the indices which label the different generations, as it will not be important for the discussion below. One may take $N$ as the lightest of the right-handed neutrinos, which is expected to dominate the production in both sectors.    
We will consider two distinct cases in which the VEV of $\phi$ is either vanishing or not.  For $\langle\phi\rangle \ne 0$, the DM mixes with the neutrinos and can be thought of as a sterile neutrino.

Throughout this paper we will discuss three possible scenarios for the dark sector internal dynamics. In the first case, the DM is a Feebly Interacting Massive Particle (FIMP), meaning the dark sector is very weakly interacting,  
 and never reaches thermal or chemical equilibrium. The second case is that of a thermalized dark sector, in which dark sector interactions drives thermalization, but to a temperature which is generally lower than the SM temperature. The dark matter freezes out while relativistic.  The third case is similar to the second, however the thermal DM is a cold relic, that is to say, it decouples from the dark sector when it is already non-relativistic.  
The cosmological constraints on these scenarios are discussed.  

The paper is organized as follows. In Sec.~\ref{sec2} we explore the possible thermal histories of the dark sector in the three different scenarios, and obtain the corresponding DM mass range that matches the correct relic abundance.
In Sec.~\ref{sec3} we consider the case where the dark sector scalar $\phi$ gets a non-vanishing VEV, resulting in mixing with the SM and the production of sterile neutrino DM. In Sec.~\ref{sec4} we discuss the constraints on light DM from astrophysics and cosmology. We conclude in Sec.~\ref{sec5}

\section{Dark Matter Relic Abundance}\label{sec2}

The thermal history of the dark sector can take several different paths, depending on its internal structure (i.e. symmetries and particle content), which lead to different quantitative predictions concerning the allowed range of dark matter mass:
\begin{itemize}
\item	FIMP: The dark matter is so weakly coupled that it never thermalizes, even within the dark sector.
\item HOT: The dark sector thermalizes at the leptogenesis scale but with a different temperature than the visible sector. The dark matter particle decouples from the hidden plasma while it is relativistic.
\item COLD: As before, the dark sector thermalizes, but the dark matter decouples from the hidden plasma while it is non-relativistic. 
\end{itemize}
Below we discuss quantitatively the implications for each of the above cases.

\subsubsection{FIMP}
In the FIMP case, the hidden sector consists of just $\chi$ and $\phi$, with $\lambda \ll 1$. 
The dark matter, $\chi$, is feebly-interacting, and its relic abundance is determined by the freeze-in mechanism~\cite{Hall:2009bx,Bernal:2017kxu}.
{For simplicity, we will assume that the scalar $\phi$ is massless, in which case its contribution to the dark matter density today is negligible.}.  The relevant Boltzmann equations (BEs) are
\begin{align}
\frac{\partial n_N}{\partial t} + 3 H n_N 
&=  -\Gamma_N \left\langle \frac{M_N}{E} \right\rangle  (n_N -n_N^{\rm eq})\,, \label{eq:approxN} \\
\frac{\partial n_\chi}{\partial t} + 3 H n_\chi 
&= \br_\chi \Gamma_N \left\langle \frac{M_N}{E} \right\rangle  n_N + 2 \leftrightarrow 2 \,, \label{eq:approxchi} 
\end{align}
where $\left\langle {M_N}/{E} \right\rangle$ is the thermally averaged inverse boost factor for the $N$ particles. The inverse decay process $\phi \chi \to N$ are neglected, and we have taken ${\br_\chi \equiv \br(N\to \chi\phi) \ll \br_{\rm SM} \equiv \br(N\to {\rm SM}) \simeq 1 }$ which is required phenomenologically as we discuss below. The late time $\chi$ abundance is found by integrating Eq.~(\ref{eq:approxchi}).  
 It is instructive to first omit  $2\leftrightarrow2$ transfer terms for simplicity, 
 in which case the equations can be solved analytically.   
One finds
\bea
Y_\chi(\infty) 
&=& Y_N(0) \br_\chi \left(1+ \frac{15 \pi \zeta(5)}{16 \zeta(3)}  \gamma_N\right),  \label{eq:relicfimp}
\eea 
where $Y(x)=n/s$ is the co-moving number density, $z=M_N/T$, and $Y_N(0)= {135 \zeta(3) g_N }/{(8 \pi^4 g_{*,M_N})}$. We have defined
\bea
\gamma_N \equiv \frac{\Gamma_N}{H_{M_N}}\,,
\eea
where the subscript $M_N$ refer to values evaluated at $T=M_N$. 
 Throughout this paper we take $\gamma_N \sim 10-100$, which corresponds to the condition for successful thermal leptogenesis~\cite{Falkowski:2011xh}.
With the above, and taking $g_{*,M_N}=106.75$,  the corresponding relic abundance is
\beq
\Omega_{\chi}^{\rm FIMP} h^2
\simeq 0.12 \times \left( \frac{m_\chi}{ \rm keV} \right)
\left( \frac{\br_\chi}{10^{-3}} \right)
\left( \frac{ \gamma_N}{25} \right) .
\eeq

\begin{figure*}
\begin{minipage}{0.49\linewidth}
\includegraphics[width=1\linewidth]{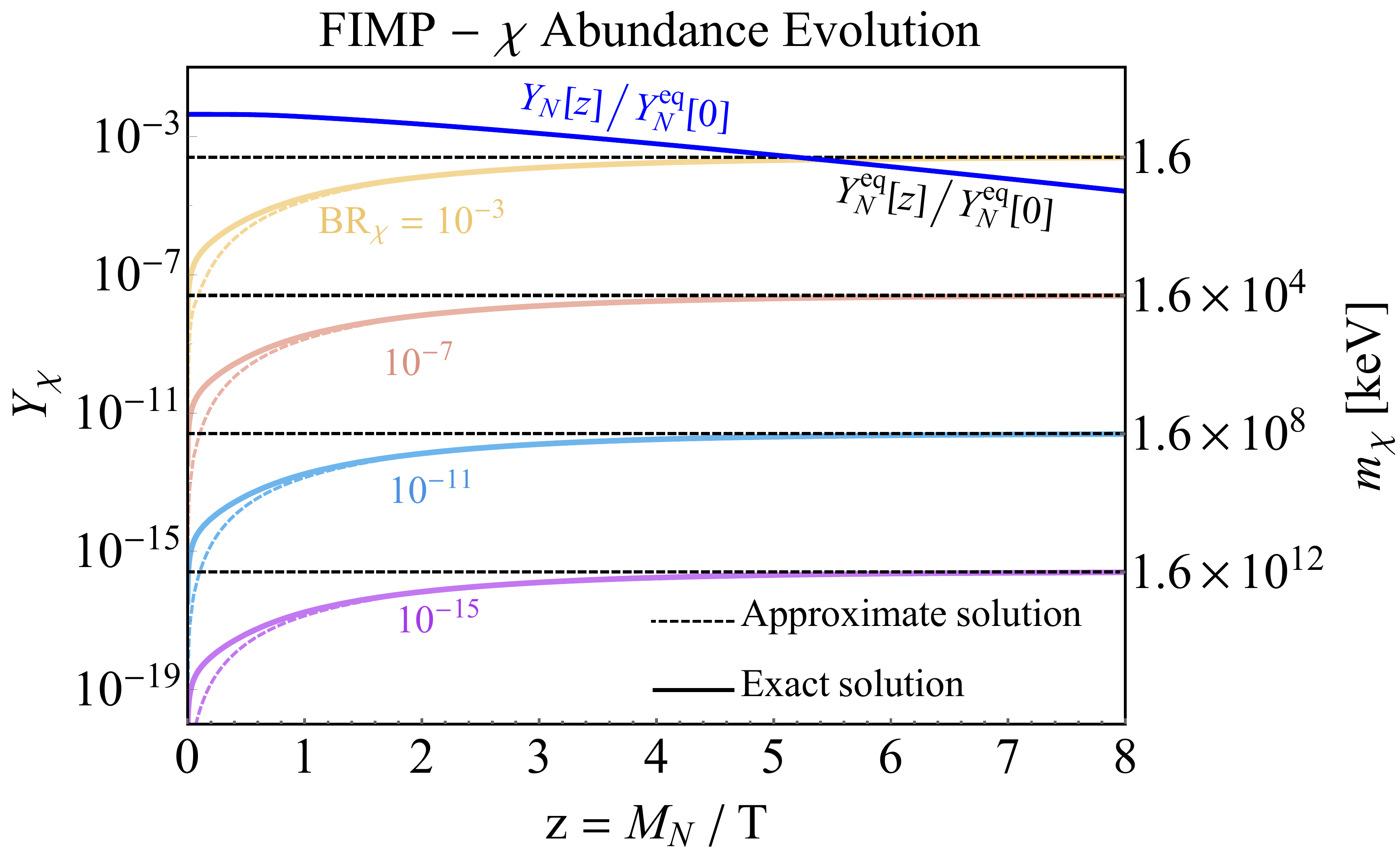}
\end{minipage}
\hfill
\begin{minipage}{0.495\linewidth}
\includegraphics[width=1\linewidth]{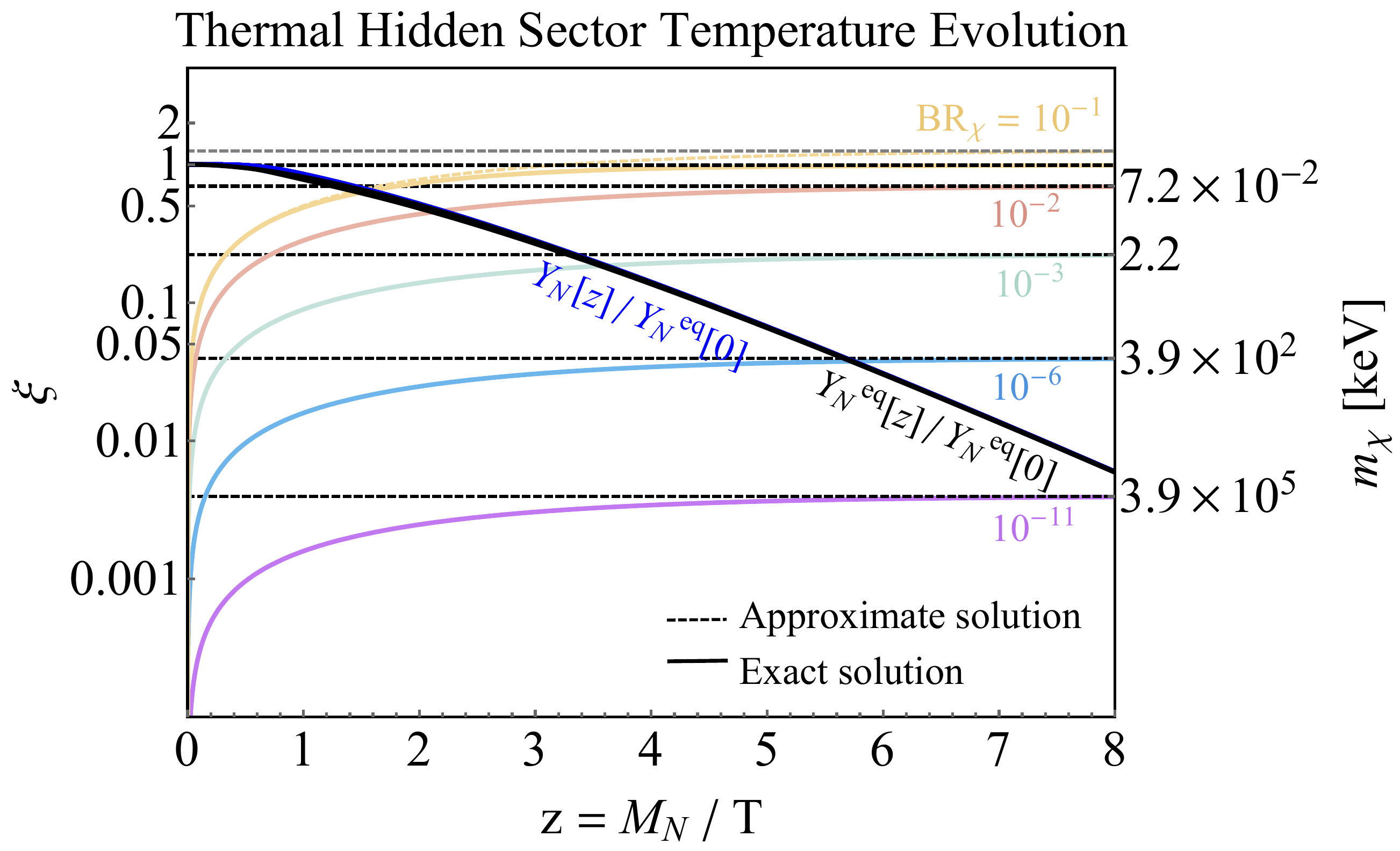}
\end{minipage}
\caption{ \label{fig:BEsol}  Solutions to the Boltzmann Equations for the FIMP~({\bf left}) and thermalized~({\bf right}) DM scenarios.
Here, $\br_{\chi} \in \left[10^{-15}, 10^{-1} \right]$, $\br_l=1-\br_\chi$, $\gamma_N = 25 $ and $g'_{*,\rm rh}=7.5$ (4 fermionic DOFs and 2 bosonic). 
The evolution of the relevant quantity is shown for each scenario with the yield of $\chi$ for the FIMP case, and the hidden to SM temperature ratio $\xi$ in the thermal case, shown on the left y-axis. On the right y-axis we show the DM mass for which the correct relic abundance is obtained. 
Different colored curves correspond to different $\br_\chi$. The solid lines are exact solutions while the dashed lines are the respective approximate solutions. For the FIMP case, the approximate solutions accurately estimate the relic abundance, and the transfer terms are negligable at large $z$. 
In the thermalized case, the approximate solutions coincide well with the exact solution for $\br_\chi < 10^{-2}$ and the backreaction from the hidden sector is negligible. For $\br_\chi\gtrsim 10^{-2}$ the backreaction becomes important, and the approximate solution overestimates the temperature ratio.} 
\end{figure*}

The derivation of the  full Boltzmann equations for this scenario are given in the appendix and summarized in  Eqs.~\eqref{eq:BEfimpN} and \eqref{eq:BEfimpchi}.  The numerical solutions for the approximate (excluding~$2\to2$ interactions)  and exact equations are shown on the left of Fig.~\ref{fig:BEsol}.  The approximate calculation for the dark matter relic abundance  agrees well with the exact solutions.

\subsubsection{Hot Relic}
In contrast to the FIMP scenario, in the thermalized sector case we assume that the hidden sector contains additional degrees of freedom, apart from $\chi$ and $\phi$, which have fast, number changing interactions with $\chi$, active during leptogenesis.
These interactions maintain chemical equilibrium in the hidden sector, causing it to thermalize at a different temperature (denoted as $T'$) than the visible sector.   
If the dark matter decouples from the hidden bath while relativistic, the current co-moving number density is the equilibrium value at that time,  
\begin{align} \label{eq:relichot}
 Y_{\chi}^{\rm}(z_f)=\frac{135 \zeta(3)}{8\pi^4 }\frac{g_{\chi}}{g_{*, \dec}}\frac{T_{\dec}'^3}{T_{\dec}^3}
 =\frac{135 \zeta(3)}{8\pi^4 }\frac{g_{\chi}}{g^\prime_{*,\rm dec}} \frac{g^\prime_{*,\rm rh}}{g_{*, \rm rh}}\frac{T_{\rm rh}'^3}{T_{\rm rh}^3}.
 \end{align}
Here the subscript ``dec" denotes the point at which $\chi$ decouples from the hidden plasma and the subscript ``rh" denotes the end of leptogenesis, i.e. reheating.
In Eq.~(\ref{eq:relichot}) we used entropy conservation from the time after reheating, to relate the ratio of temperatures at decoupling to the ratio after reheating. 
Here, and throughout the paper, we do not distinguish between entropy and energy degrees of freedom. 

Excluding inverse decays,
the Boltzmann equation for $N$ is the same as in the FIMP scenario, and is given by Eq.~(\ref{eq:approxN}). Here, the hidden sector evolution is best studied in terms of the energy density, 
\beq \label{eq:approxhot}
\frac{\partial \rho_d}{\partial t} + 4 H \rho_d 
= \br_\chi \Gamma_N M_N n_N + \gamma_{lh \to \chi \phi}\,,
\eeq
where $\rho_d=\rho_\chi+\rho_\phi+\rho_{\rm hidden}$ is  the total energy density in the hidden sector, with $\rho_{\rm hidden}$ serving as a general name for the states in the sector which are not $\chi$ and $\phi$. 
$\gamma_{lh \to \chi \phi}\equiv~\frac{2}{\pi^3} \br_\chi \br_l \Gamma_N  M_N^4 \frac{I'(z)}{z}$ is the thermally averaged rate of energy transfer from the SM to the hidden sector, and $I'(z)$ transfer integral is defined in the appendix, Eq.~(\ref{eq:Iprimedef}).  
We consider $T'\ll T$, so inverse decays and energy transfer from the hidden sector to the SM can be neglected.  
This approximation fails at $T' \simeq T$, when the two sectors thermalize and the correct solution requires solving the full BEs.

Assuming the dark sector is always thermalized, so that the  energy density takes on its equilibrium form $\rho_d= \rho_d^{\rm eq}(z^\prime) = \frac{\pi^2 }{30}\frac{g'_{*, \rm rh}}{ z'^4}M_N^4$, Eq.~(\ref{eq:approxhot}) 
 can be integrated and solved for the temperature ratio $\xi(z)\equiv z/z'=T'/T$.  One finds,
\begin{align}
\xi (z)
&=\gamma_N \br_\chi \left[ \left( \frac{4}{3}\frac{g_{*}}{g'_{*,\rm rh}} \int_0^{z} dz  z^2 Y_N \right. \right. \nn \\ 
&+\left. \left. \frac{180}{\pi^4} \frac{1}{g'_{*,\rm rh}} \br_l \int_0^z dz z^4 K_2(z) \right) \right]^{1/4}\,.
\end{align}
Under the assumption that $N$ is in thermal equilibrium with the SM sector,  the temperature ratio of the two sectors immediately proceeding decoupling is  
\beq
\xi_{\rm rh}^4\to {\rm Min}\Bigg[ \frac{240}{\pi^4} \frac{ \gamma_N \br_{\chi} }{g'_{*,\rm rh}} \left( g_N  + 12\br_l \right),1\Bigg].
\eeq 
Inserting this result back into Eq.~(\ref{eq:relichot}) we find the late times yield for $\chi$
\beq
Y_{\chi}(\infty) \approx 0.41 \times \frac{g_{\chi}}{g_{*,\rm rh}}\frac{g^\prime_{*,\rm rh}}{g^\prime_{*,\dec}}\left(\frac{ \gamma_N \br_{\chi} }{g^\prime_{*,\rm rh}} \left(g_N+12\br_l\right) \right)^{3/4}, 
\eeq
and the corresponding relic abundance 
\begin{align}
\Omega_\chi^{\rm{hot}} h^2  
	&\simeq 0.12 \times \left( \frac{m_\chi}{0.15~\rm keV} \right) \nn  \\
	 &\times
    \left[ \left(\frac{ \br_{\chi} }{10^{-3}  } \right)
    \left( \frac{\gamma_N}{25} \right) 
    \left( \frac{g_N + 12\br_l}{14} \right)  \right]^{3/4},
\end{align}
where we have taken $\chi$ to be the lightest dark sector particle and thus $g^\prime_{*,\rm dec}=\frac{7}{8}g_{\chi}$.

For the same right-handed neutrino decay parameters, the relic abundance for the thermalized case is greater than for the FIMP case. When the right-handed neutrino decays, it produces relativistic $\chi$ particles with large amounts of kinetic energy. In the FIMP case, the number of particles is fixed from these decays. However, if the dark-sector thermalizes, the $\chi$ particles efficiently convert this excess kinetic energy to additional particles (of both $\chi's$ and other hidden sector states). This results in a larger number of $\chi$ particles in the thermalized case, and ultimately, a larger relic abundance.

The full Boltzmann equations for this scenario are given in the appendix, Eqs.~\eqref{eq:hotN}~\eqref{eq:hotdark}, and the numerical solutions for the approximate and exact equations are shown on the right of Fig.~\ref{fig:BEsol}.
We find the approximate solutions, which include energy transfer terms from the SM to the hidden sector to trace the exact solution well.  
The hidden sector back-reaction as well as the inverse decays from it, can be completely neglected.
However, if the decay rate to the hidden sector becomes too large, the visible and hidden sectors begin to thermalize ($\xi \to 1$) and back-reaction terms can no longer be neglected. 
For this region of parameter space, the approximate solution overestimates the hidden sector temperature, as shown in the top-most curve on the right of  Fig.~\ref{fig:BEsol}.

\begin{figure*} 
\center
\includegraphics[width=1\linewidth]{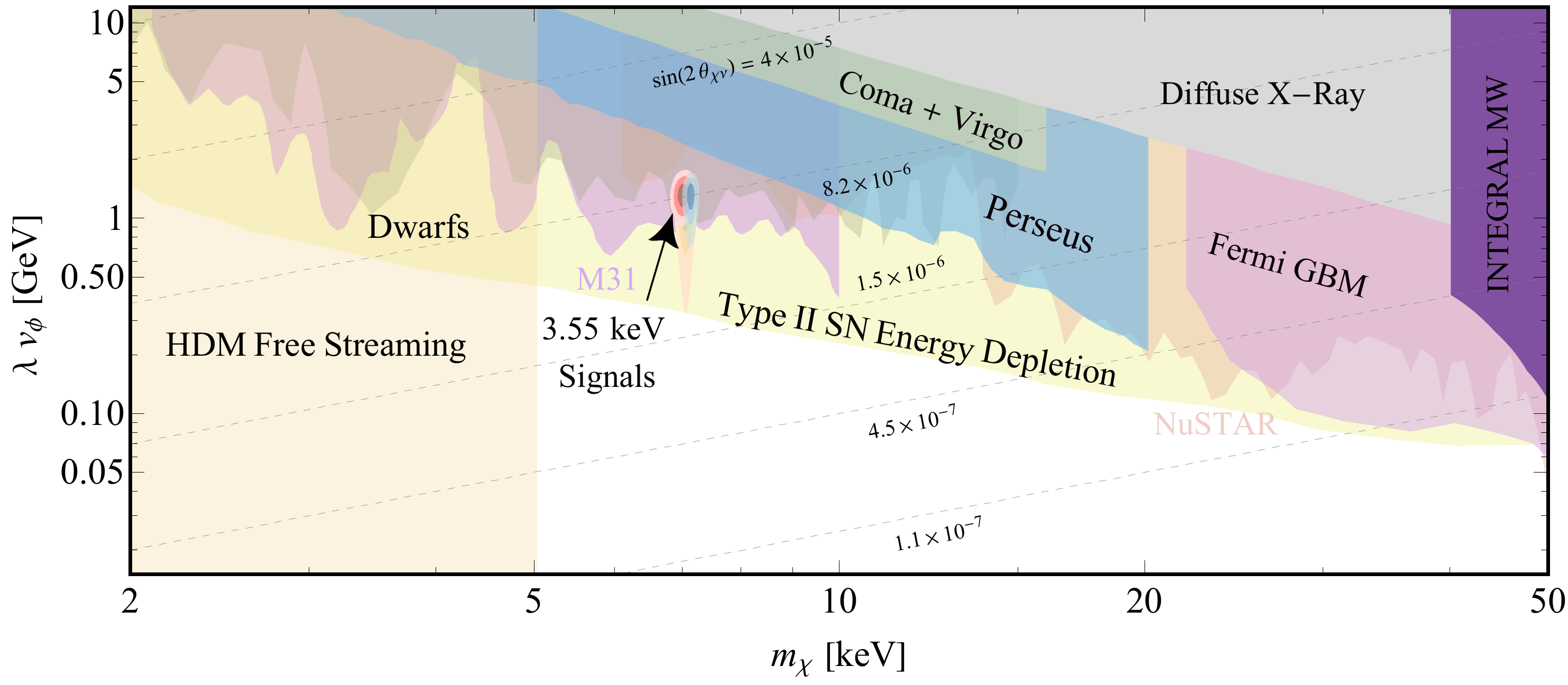} 
\caption{ \label{fig:sNDM} Bounds on the mass and mixing parameter of sterile neutrino dark matter, as well as the best-fit region for the indirect detections in X-ray measurements, adopted from~\cite{Abazajian:2017tcc}. 
Constraints are obtained from several X-ray observations~\cite{Boyarsky:2005us,Boyarsky:2006zi, Boyarsky:2007ge, Horiuchi:2013noa, Malyshev:2014xqa, Ng:2015gfa}, as well as Type II supernova core energy depletion due to sterile-active neutrino mixing enhanced by the MSW effect~\cite{Arguelles:2016uwb}.
The $3.55~\rm keV$ detection signal dark$\to$light shaded regions correspond to $1,2$ and $3\sigma$ detection in the various clusters ({M31, PN and MOS Clusters}~\cite{Bulbul:2014sua,Boyarsky:2014jta}).  The masses shown correspond to hot and FIMP DM production scenarios, with $\br_\chi \sim 10^{-5},10^{-4}$ for the hot and FIMP cases, respectively, while a constant $\gamma_N=25$ is used. Diagonal dashed lines represent constant sin$2\theta_{\chi \nu}$. 
The vertical orange shaded region refers to the free streaming bound for the hot DM scenario.
The analogous bound in the FIMP case is weaker, such that all $m_\chi$ in the displayed range are allowed.}
\end{figure*} 

\subsubsection{Cold Relic}
In the third case, the dark sector follows the same thermal evolution path as in the previous section, until it reaches $\xi_{\rm rh}$. 
However, at some point after the dark and SM sectors decouple, $\chi$ freezes out, i.e.,  it decouples from the hidden plasma when it is non-relativistic.  
We assume that the dark sector contains additional degrees of freedom lighter than $m_\chi$ into which DM can annihilate.   
Then the BE for $\chi$ can be written in the standard freeze-out language
\bea \label{eq:cold}
\frac{\partial n_\chi}{\partial t} + 3 H n_\chi &=& -\gamma_{ann}'\left( \frac{n_{\chi}^2}{n_{\chi,eq}'^{2}} - 1  \right) ,
\eea
where $\gamma_{ann}' = \langle {\sigma v} \rangle' n_{\chi,eq}'^2$ is the thermally averaged annihilation rate, noting that for this case the hidden sector begins the freeze-out process at a different temperature than the SM. 

The above can be solved approximately using the usual sudden freeze-out approximation.  One finds, 
\begin{equation}
\Omega_\chi   = \frac{s_0}{\rho_c }  \frac{\sqrt{g_{* , m }}}{\sqrt{g_{* , f }}}\frac{m_\chi H_m }{s_m } \frac{x_f^\prime}{ \langle  \sigma v \rangle_f} \xi_{\rm f}\,,
\label{eq:coldx}
\end{equation}
where $x^\prime=m_{\chi}/T^\prime$,  $s_0$ is the entropy density of the  SM bath today, $\rho_c$ is the critical density, and the subscripts $m$ and $f$ denote quantities at temperatures $T=m_\chi$ and freeze-out, respectively. 
Eq.~(\ref{eq:coldx}) is very similar to the usual freeze-out equation, however it includes an additional factor $\xi_{\rm f} < 1$, which implies a suppressed annihilation cross-section, $\langle \sigma v \rangle'$, compared to the more common freeze-out case  with $\xi_{\rm f} =1$.
This suppression implies a more weakly coupled particle than the standard WIMP scenarios, and therefore disfavors indirect-detection.

\section{Sterile Neutrino Dark Matter}\label{sec3}

Let us now consider the model described by Eq.~\eqref{lag}, but for the case where the $\phi$ field acquires a non vanishing VEV, $\langle \phi \rangle = v_\phi \neq 0$. 
Upon integrating out the heavy right-handed neutrinos, a small Majorana mass for $\chi$,   $\tilde{m}_\chi=m_\chi\pm\frac{(\lambda v_\phi)^2}{2M_N}$, is induced alongside a mixing between 
the SM neutrinos and the DM candidate \cite{Falkowski:2011xh}.  One finds the mixing angle
\beq
\sin\theta_{\chi \nu} 
\simeq \frac{\lambda v_{\Phi}   }{ m_\chi} \sqrt  \frac{m_\nu}{M_N},
\eeq
where $m_\nu$ is the physical neutrino mass. 
Consequently, $\chi$ can be viewed as a sterile neutrino.

This mixing naturally induces an irreducible Dodelson-Widrow (DW) contribution to the DM density, via the oscillations of the active neutrino~\cite{Dodelson:1993je, Kusenko:2009up}.
For the relevant region of parameter space discussed below, the effects of the $\chi\leftrightarrow \nu$ oscillations on the relic abundance is typically smaller than $\mathcal{O}(10^{-3})$ and can be readily neglected. 

The mixing allows for several new decay modes for DM. Firstly, if $\chi$ is lighter than the
electroweak gauge bosons, then it may decay via a 3-body process through an off-shell $Z^*/W^*$, $\chi \to \nu(Z^*\to f\bar{f})$ and $\chi \to e^-(W^*\to f\bar{f}')$, where $f/f'$ are SM fermions.
Secondly, for $\chi$ lighter than MeV, the main observable process will be a 2-body decay, induced at one loop, to a neutrino and a monochromatic photon, $\chi \to \nu \gamma$.
The relevant decay widths are given by~\cite{Abazajian:2001vt,Pal:1981rm}
\begin{align}
\Gamma_{\chi \to 3\nu} 
&= \frac{1}{768\pi}\frac{\alpha^{2}}{s_{w}^{4}m_{W}^{4}}s_{2\theta}^{2}m_{\chi}^{5} \nn \\
&\simeq  1.7 \times 10^{-25} {\rm s}^{-1} \left(\frac{s_{2\theta} }{10^{-5}} \right)^2 \left( \frac{m_\chi}{10~\rm keV} \right)^5, \\
\Gamma_{\chi \to \nu  \gamma} 
&=\frac{9}{2048\pi^{2}}\frac{\alpha^{3}}{s_{w}^{4}m_{W}^{4}}s^{2}_{2\theta} m_{\chi}^{5} \nn \\
&\simeq  1.38 \times 10^{-27} {\rm s}^{-1}\left(\frac{s_{2\theta} }{10^{-5}} \right)^2 \left( \frac{m_\chi}{10~\rm keV} \right)^5 .
\end{align}
The rates for these processes cannot be arbitrarily large, and are constrained by DM stability and x-ray and gamma-ray measurements. 
Dark matter stability requires a lifetime of $\tau_{\rm DM} \gtrsim 5\times 10^{18}~\rm sec$~\cite{Audren:2014bca},  
 which limits the neutrino mixing angle mainly from the $\chi \to 3\nu$ decay. 
For $m_\chi \simeq 10 \rm~keV$, the bound can be written in terms of 
$\lambda v_\phi$,
\begin{align}
\lambda  v_{\Phi}  
\lesssim& ~1.6 ~\rm{TeV}\nn \\
\times&    \left( \frac{0.05 \ev }{m_{\nu}} \right)^{1/2}  \left( \frac{M_N}{10^{11} \gev}  \right)^{1/2} 
\left( \frac{10~\rm keV}{m_\chi}  \right)^{3/2}, 
\end{align}
 which is trivially met for $v_\phi\lesssim v_{EW}$, $\lambda \lesssim 1$. 
 Another constraint comes from consistency with cosmic ray and diffuse gamma observations, which requires a decay time 
 $\tau_{\rm DM}\gtrsim 10^{26}~\rm sec$ \cite{Essig:2013goa,Abazajian:2017tcc}, 
thus setting a bound on the neutrino mixing angle from the $\chi \to \nu \gamma$ decay at $\lambda  v_{\Phi} \lesssim ~0.015-12 ~\rm{GeV}$ for $m_\chi\sim 2-50~\rm keV$.   In Fig.~\ref{fig:sNDM}, we compile these constraints from various X-ray observations: M31 Horiuchi et al.~\cite{Horiuchi:2013noa}, stacked dwarfs~\cite{Malyshev:2014xqa}, the diffuse X-ray background~\cite{Boyarsky:2005us}, individual clusters “Coma+Virgo”~\cite{Boyarsky:2006zi}, Fermi GBM~\cite{Ng:2015gfa} and INTEGRAL~\cite{Boyarsky:2007ge}. 
We further show the constraint from Type II supernova core energy depletion due to sterile-active neutrino mixing enhanced by the MSW effect~\cite{Arguelles:2016uwb}.  The free streaming limit to be discussed below, and relevant only for the hot DM case, is also shown.
\begin{figure}
\includegraphics[width=1
\linewidth]{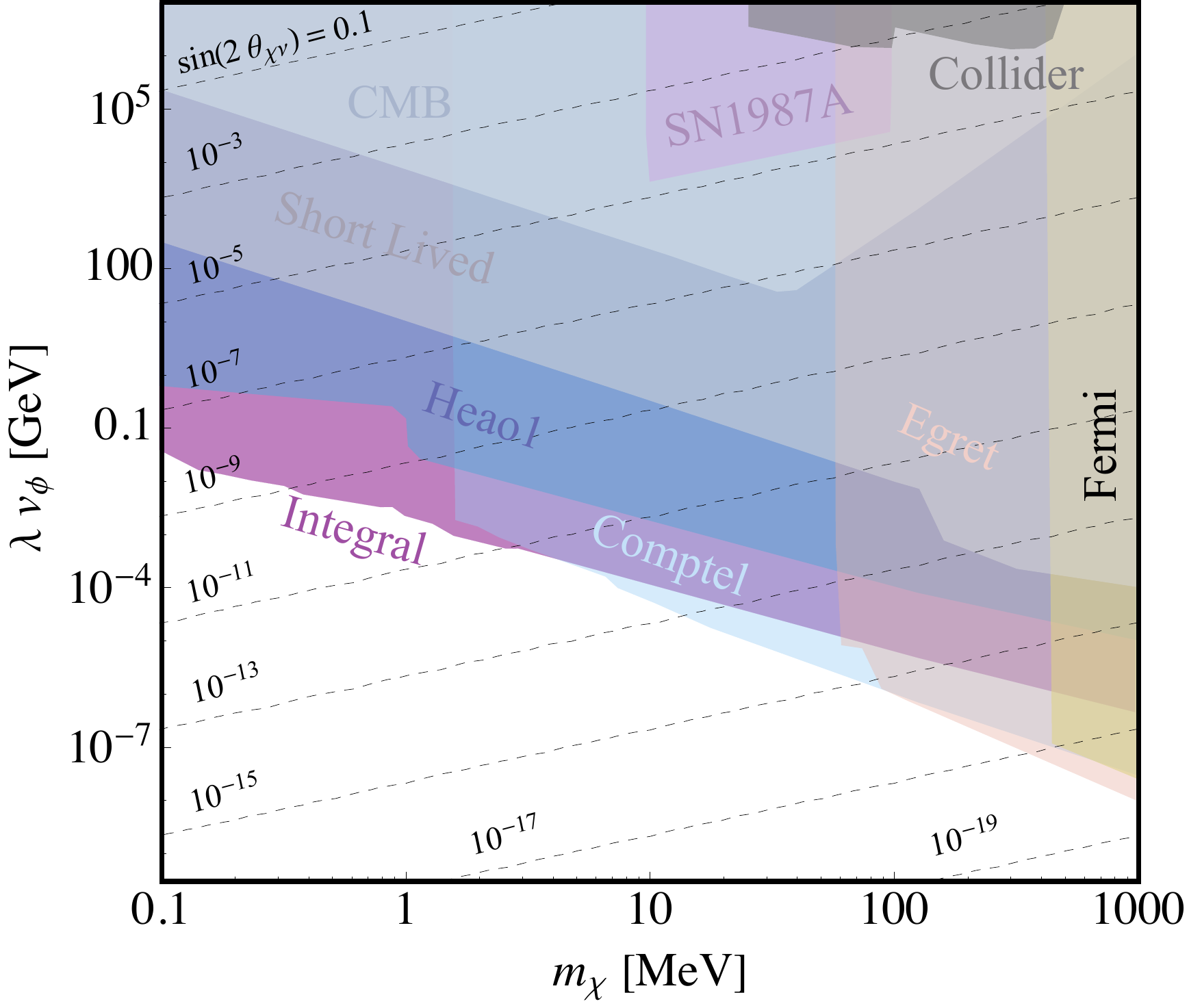}
\caption{ \label{fig:sNDM2} Bounds on the mass and mixing parameter of sterile neutrino dark matter, for MeV-GeV DM masses. 
Constraints are obtained from a combination of X-ray observations (Integral, Heao1, Comptel, Egret and Fermi LAT), CMB, SN1987A cooling and collider searches. Within the light gray region, the DM lifetime is shorter than the age of the Universe. These constraints are summarized in~\cite{Smirnov:2006bu, Essig:2013goa,Shakya:2015xnx, Berlin:2016bdv}.
The masses shown correspond to hot and FIMP DM production scenarios, with $\br_\chi \sim 10^{-11}-10^{-6},10^{-9}-10^{-5}$ for the hot and FIMP cases, respectively, while a constant $\gamma_N=25$ is used. Diagonal dashed lines represent constant sin$2\theta_{\chi \nu}$. }
\end{figure}
On top of these constraints, Fig.~\ref{fig:sNDM} shows the predicted mass for a FIMP and relativistically-decoupled sterile neutrino dark matter.  These dark matter masses correspond to  heavy right-handed neutrino branching fractions of order $10^{-4}$ and $10^{-5}$ respectively.

In Fig.~\ref{fig:sNDM2},  we show the constraints on $100$~keV to $1$~GeV DM masses, corresponding to to right-handed neutrino branching fractions spanning ($10^{-11}-10^{-5}$).
Between $0.1-1$~MeV, the relevant bounds are mainly from X-ray observations performed by Heao1~\cite{Gruber:1999yr} and Integral~\cite{Bouchet:2008rp}
, as well as measurements of the CMB power spectrum~\cite{Smirnov:2006bu}. Beyond $1$~MeV, processes involving charged fermions in the final state must be included, for which the width is computed in \cite{Essig:2013goa} as
\begin{align}
	\Gamma_{\chi \to \nu e^+e^-} 
	&= 
	\frac{0.59}{192\pi}\frac{\alpha^{2}}{s_{w}^{4}m_{W}^{4}}s_{2\theta}^{2}m_{\chi}^{5} \nn \\
	&\simeq  4.2 \times 10^{-10} {\rm s}^{-1} \left(\frac{s_{2\theta} }{10^{-5}} \right)^2 \left( \frac{m_\chi}{10~\rm MeV} \right)^5,
\end{align}
and an additional bound from Comptel observations~\cite{Comptel} is included.
Above $10$~MeV, additional bounds from SN1987A cooling~\cite{Kainulainen:1990bn} and direct detection searches for neutral leptons~\cite{Bernardi:1985ny, Bergsma:1985is, Bernardi:1987ek,Vaitaitis:1999wq, PIENU:2011aa,Ruchayskiy:2011aa} are relevant. For $100$~MeV and above, the constraints come from Gamma ray observations by Fermi LAT~\cite{Ackermann:2012pya} and Egret~\cite{Strong:2004de}.
We conclude that constraints on the $\nu/\chi$ mixing are fairly weak, and are easily accommodated within the parameter space of our framework, while still providing a possible signal for existing and future observations.

The potential detection of sterile neutrino DM has gained much interest in the past several years, due to an unidentified photon emission line at $E_\gamma\simeq 3.55$~keV in X-ray measurements from cluster observations ~\cite{Bulbul:2014sua, Boyarsky:2014jta}.
A summary of the most recent data analyses regarding the signal from several sources ({Chandra, Hitomi, XMM-Newton, Suzaku}), as well as a sterile neutrino DM signal fit, is found in \cite{Abazajian:2017tcc} and shown in Fig.~\ref{fig:sNDM}. Note, that due to disagreement in the literature and systematic uncertainties, both the signals and the constraints should be taken with a grain of salt. 
While the best-fit region for the sterile-neutrino scenario is disfavored by the supernova constraints as well as by some of the X-ray measurements, we point out that the scenario proposed in this section provides a natural explanation for the alleged emission line, since the required mass range and mixing angle to match the $3.55$~keV energy are naturally obtained in our model, with $m_\chi =~7.1~ \rm keV$,~$\lambda v_\phi \simeq 1.3~ \rm GeV$.

\begin{figure*}
\begin{minipage}{0.49\linewidth}
\includegraphics[width=1\linewidth]{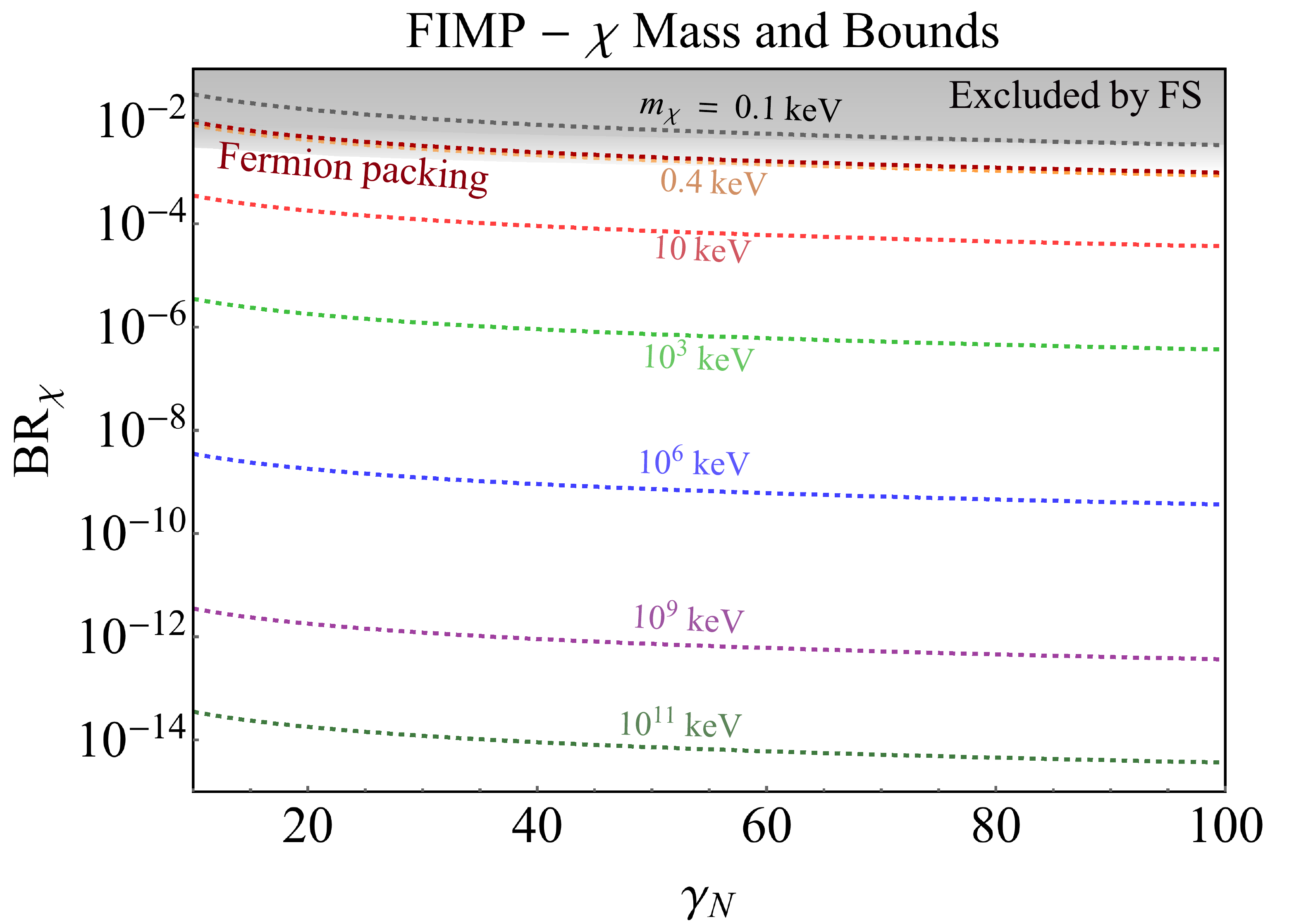}
\end{minipage}\hfill
\begin{minipage}{0.49\linewidth}
\includegraphics[width=1\linewidth]{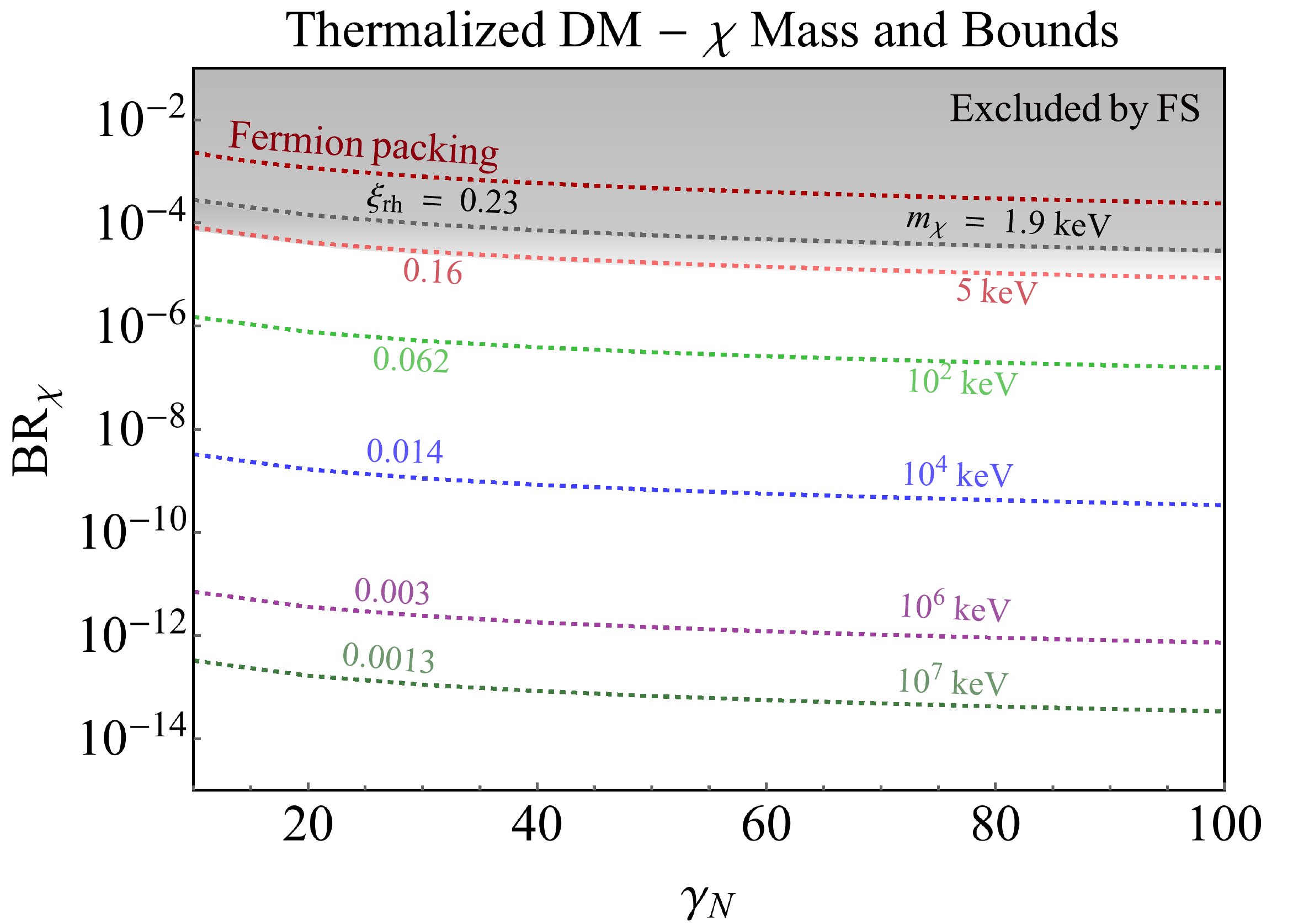}
\end{minipage}
\caption{\label{fig:FSsol} 
Contour plots of $m_\chi$ for the FIMP~({\bf left}) and hot ({\bf right}) DM scenarios, showing our full parameter space in the $\br_\chi-\gamma_N$ plane. Here, $\br_{\chi} \in \left[10^{-15}, 10^{-2} \right]$, $\br_l=1-\br_\chi$, $\gamma_N \in \left[10, 100\right]$.
The gray region represents the excluded region of DM masses  where $\lambda_{\rm{FS}} > [0.05, 0.15]~\rm Mpc$, where the darker area corresponds to the more conservative constraint, and the lighter shading to the more stringent values, taking into account uncertainties in the DM density power spectrum~\cite{Garzilli:2015iwa}. Further qunatitative discussions on free streaming in this paper employ the stringent constraint.
The  dashed thick red line is the fermionic phase-space bound. 
For the FIMP case, there exists a lowest possible DM mass, $m_\chi \gtrsim 0.41~ \rm keV$ for $\gamma_N=100, \br_\chi\simeq 10^{-3}$. In the hot DM case, the bound on the DM mass is a global one. It is set at $m_\chi \gtrsim 5~ \rm keV, \xi_{\rm rh}\leq 0.16$, and can be obtained by a range of $\br_\chi, \gamma_N$. }
\end{figure*}

\section{Cosmological Constraints}\label{sec4}

Finally, we discuss the main cosmological constraints that often arise when considering light DM, namely, the free streaming limit, Big Bang Nucleosynthesis (BBN) and the phase-space, or ``packing", bound on fermionic DM.

\subsubsection{Free Streaming}

A stringent constraint on the light DM mass comes from its  free streaming length. The free streaming of dark matter particles, due to a non-trivial velocity dispersion, erases matter density perturbations, and therefore structure formation, on scales $\lambda<\lambda_{\rm{FS}}$. Consequently, the most stringent bound on $\lambda_{\rm{FS}}$ arises from small structure formation $\lambda_{\rm{FS}}<\mathcal{O}(0.1)~\rm Mpc$~\cite{Boyarsky:2008xj,
Sigurdson:2009uz,
Das:2010ts, Viel:2013apy, Berlin:2017ftj}.
The free streaming length is defined by the particle horizon of the DM~\cite{Kolb:1990vq}\footnote{It is shown in~\cite{Das:2010ts}
 that this approximation for $\lambda_{FS}$ agrees with a derivation based on a momentum averaged velocity dispersion up to the few percent level.}
\beq\label{eq:genFS}
\lambda_{\rm{FS}} \equiv
\int_{a_{\rm rh}} ^{a_{\rm eq}}\frac{\langle v(a) \rangle}{a^2 H}da 
\simeq \frac{a_{\rm{NR}}}{H_0 \sqrt{   \Omega_R}}
\left( 0.62
+  {\rm ln}\Bigg[\frac{a_{\rm eq}}{a_{\rm{NR}}} \Bigg]  \right), 
\eeq
where $a(t)$ is the FRW scale factor, normalized so that today's value is $a_0$, $v(a)$ is the average DM velocity at a given time, $a_{\rm rh}$, $a_{\rm eq}$ are the scale factors at reheating and at matter/radiation equality, respectively, and $a_{\rm NR}$ is the scale factor when DM becomes non-relativistic.
To quantitatively estimate $\lambda_{\rm FS}$, we use the results obtained by Planck~\cite{Ade:2015xua}
 for the cosmological parameters~$H_0=67.3~{\rm km~ s^{-1}~ Mpc^{-1}}$,~$\Omega_R=9.3\times 10^{-5}$,~and $a_{\rm eq}=2.9\times10^{-4}$.

Evaluation of the free streaming length strongly depends on the production and decoupling mechanism of DM
~\cite{Das:2010ts,
Sigurdson:2009uz,Boyarsky:2008xj}, and is therefore different for each of our relativistic decoupling scenarios (hot, FIMP), and must be considered separately.
In the FIMP scenario, the DM particles are produced by $N$ decays, thus with initial relativistic momentum $p_{\rm rh}=E_{\rm rh}=M_N/2$. Using the instantaneous decay approximation $H(T_{\rm rh})=\Gamma_N$ and defining the non relativistic momentum as $p_{\rm NR}=m_{\chi}$, the non-relativistic scale factor is
\begin{align}
	 a_{\rm NR}^{\rm FIMP} 
	= \frac{ T_0}{2 m_\chi}  \left( \frac{g_{*s,0}} {g_{*s,\rm rh}} \right)^{1/3} \gamma_N^{-1/2} .
\end{align}
Taking $g_{*,0}=3.91$, $g_{*,\rm rh}=106.75$, and $T_0=2.35\times10^{-4}~\rm eV$, 
\begin{align}
\label{eq:lfsFIMP}
	\frac{\lambda_{\rm FS}^{\rm FIMP}}{\rm Mpc}  
	&\simeq 
	2.8 \times 10^{-2}  \left(\frac{\rm keV}{ m_\chi}\right)\left(\frac{50} { \gamma_N}\right)^{1/2} \nn \\
	&\times
	\left( 1+ 0.09~{\rm ln}
	\left[\left(\frac{m_\chi}{\rm keV}\right)\left(\frac{\gamma_N} {50}\right)^{1/2}\right]
	 \right).
\end{align} 

Next, we consider the thermalized hot DM scenario. Here, the DM velocity distribution is governed by the number of relativistic degrees of freedom, and the evolution of $T'/T$. 
By utilizing entropy conservation separately in each sector,  the non relativistic scale factor is found to be 
\begin{align}
a^{\rm{Hot}}_{\rm{NR}}=\left(\frac{g_{*,0}}{g_{*,\rm rh}}\frac{g'_{*,\rm rh}}{g'_{*,\rm NR}} \right)^{1/3} 
\frac{3.15 T_0}{m_{\chi}}\xi_{\rm rh}
= 1.3\times 10^{-5} \xi_{\rm rh}^4,	
\end{align}
where $g_{*,\rm rh}=106.75,g'_{*,\rm rh}=7.5,g'_{*,\rm NR}=3.5$, and we have used Eq.~(\ref{eq:relichot}) to relate the mass of $\chi$ to the reheating temperature. 
 Plugging this result in Eq.~\eqref{eq:genFS}, the free streaming length for hot DM is given by
\begin{align}
\label{eq:lfsHOT}
	\frac{\lambda_{\rm{FS}}^{{\rm Hot}}}{\rm Mpc}
&\simeq
7.8 \times 10^{-3}  \left( \frac{\xi_{\rm rh}}{0.1} \right)^4~
\left( 1 + 0.31~{\rm ln} \left[ \frac{0.1}{\xi_{\rm rh}} \right]\right) .
\end{align}

A more detailed derivation for the free streaming scale in these scenarios, as well as a calculation for a relic which was, at early times, in thermal equilibrium with the SM, is given in Appendix B.
{Note that the different cosmological history in the hot DM and FIMP cases leads to a completely different parametric dependence of   $\lambda_{\rm{FS}}$, cf. Eq.~\eqref{eq:lfsFIMP} 
and  Eq.~\eqref{eq:lfsHOT}.  
}

In Fig.~\ref{fig:FSsol}, we show the $\chi$-mass contours on the Br$_\chi$-$\gamma_N$ plane, for which the correct relic abundance is obtained with the full BEs.   
Regions excluded by free streaming are shown in gray. For the FIMP case,  scanning the parameters reveals a lowest bound on $m_\chi \gtrsim m_\chi^{\rm FIMP} = 0.41~ \rm keV$ for $\gamma_N=100$ and $\br_\chi\simeq 10^{-3}$, at the Fermion Packing bound, whereas for the hot case, the bound is $m_\chi \gtrsim m_\chi^{\rm hot} = 5~ \rm keV$.  In both cases, light (order keV) DM is indeed allowed within this framework.

\subsubsection{Big Bang Nucleosynthesis}
We now  consider the bound on relativistic DM imposed by BBN.
The formation of light elements during BBN constrains the number of relativistic degrees of freedom in the early universe. 
This constraint is quantified by the effective number of relativistic neutrinos~$N_{\nu,eff}=3.15 \pm 0.23$~\cite{Ade:2015xua}. 
Any contributions of relativistic DM candidates at BBN must be bound by $N_{\nu,\rm eff}$, which can be parametrized as
~\cite{Csaki:2017spo}
\beq
N_{\nu,\rm eff}
= N_{\nu,\rm eff}^{\rm{SM}} + \frac{4}{7} \left( \frac{11}{4} \right)^{4/3} g'_{*,\rm eff}\,.
\eeq
Here the contribution predicted by the SM is $N_{\nu,\rm eff}^{SM}=3.046$ and 
 $g'_{*,\rm eff}=\sum_i g'_{i} s_i \left( \frac{T'_i}{T} \right)^4 $ is the number of relativistic degrees of freedom in the hidden sector at BBN with $s_i=$1  (7/8) for a boson (fermion).
At late times, the hidden to standard temperature ratio is given by entropy conservation in each sector.  Explicitly 
 \beq \label{eq:Tratio}
 \frac{T'_{\rm{BBN}}}{T_{\rm{BBN}}}
 =\left( \frac{g_{*,{\rm{BBN}} }}{g'_{*,{\rm{BBN}}}}
 \frac{g'_{*,{\rm rh}}}{g_{*,\rm rh}} \right)^{1/3} \xi_{\rm rh}.
\eeq
 With the above, for the hot DM case one finds  $\Delta N_{\rm eff}\equiv N_{\nu,\rm eff} - N_{\nu,\rm eff}^{\rm SM} \simeq  1.75 \xi_{\rm rh}^4$, which satisfies the 2$\sigma$ bound, $\Delta N_{\rm eff}\leq0.61$, for $ \xi_{\rm rh} \lesssim 0.77$.  We conclude that the constraint from BBN is much weaker than the one from  free streaming. 

\subsubsection{Fermion packing}
Lastly, we consider the phase-space bound.
For fermionic DM candidates, a very robust lower bound on their mass can be obtained due to the Pauli exclusion
principle. 
The number of fermions that can be ``packed" in a given region of the phase space is limited. 
 Decreasing the particles mass inevitably increases their number in a given gravitationally bound object, when ``packed" in the densest possible formation.
The requirement that the phase-space density of the DM does not exceed that of the degenerate Fermi gas leads to a lower mass bound. 
For a spherically symmetric object, the bound reads~\cite{Tremaine:1979we} 
\begin{equation}
	m_{\rm DFG}^4
	\geq  \frac{9\pi}{4\sqrt{2} g M^{1/2}R^{3/2}G_N^{3/2}},
\end{equation}
where $g$ is the number of DM degrees of freedom, $M$ and $R$ are the bound system's mass and radius, respectively.
A detailed analysis is done for a number of dwarf galaxies in~\cite{Boyarsky:2008ju}, and sets the limit at $m_{\rm DFG}\geq 0.41~\rm keV$. This constraint is weaker than the free-streaming bound for the hot DM scenario, and marginally stronger for the FIMP scenario in a portion of its parameter space, as shown in Fig.~\ref{fig:FSsol}.

\section{Summary}\label{sec5}

In this paper we discussed the dynamics and phenomenological implications of an interacting hidden sector in the context of two sector leptogenesis. We find that a shared production mechanism for leptogenesis and symmetric dark matter yields a relic abundance of dark matter that depends on several main factors, namely, the branching fraction to the hidden sector, the decay rate of the heavy right-handed neutrinos, and the number of degrees of freedom in the hidden sector. Consequently, the ratio of dark matter number density to the baryon number density is rather sensitive to the model parameters, thereby accommodating a wide range
of dark matter masses. 
Nonetheless, unless decays of the right-handed neutrinos to the dark sector are extremely rare, one naturally expects light dark matter, in the keV to GeV mass range.

In the case where the hidden scalar which couples to the right-handed neutrino gets a VEV, the mechanism can be viewed as a production mechanism for decaying (and possibly interacting) sterile neutrinos.  This may have interesting implications for the dark matter interpretation of the  keV line reported by various X-ray observations.

 In this work we asserted that the DM abundance is dominated by the fermionic candidate $\chi$. It is  plausible that the DM is composed mainly of the scalar $\phi$, or some mixture of the two, depending on the mass ratio between the two particles. A configuration such as this will lead to two main differences in the resuling phenomenolgy. Firstly, the astrophysical bound coming from ``fermion packing'', discussed in section~$\RN4$.3, is alleviated, based on the amount of bosonic density compared to fermionic. Secondly, if the dark matter energy density is dominated by the scalar component, the number of produced sterile neutrino DM particles would be significantly smaller, and so of lesser phenomenological interest in the mass range we have considered in this letter. We leave the scalar dominated DM scenario for further work.

While we have focused on the concrete scenario of thermal leptogenesis, there are several possibilities and modifications that could prove interesting for further exploration.
Furthermore, we have chosen not to specify the precise nature of the interactions in the hidden sector in favor of a broader study. 
It would be interesting to consider a detailed model for interactions within this framework, such as a hidden photon that kinetically mix with the SM.  Such models  could naturally predict DM self-interactions as well as additional potential discovery channels.

\section{Acknowledgments}
We would like to thank Josh Ruderman for collaboration at early stages of this work, and Alexey Boiarskyi for useful comments. 
The work of AF is supported by the ERC Advanced Grant Higgs@LHC and by  the European Union's Horizon 2020 research and innovation programme under the Marie Sk\l{}odowska-Curie grant agreement No 690575.
EK and TV are supported in part by the I-CORE Program of the Planning Budgeting Committee and the Israel Science Foundation (grant No. 1937/12), and by the Binational Science Foundation (grant No. 2016153). The work of EK is further supported by the  Israel Science Foundation (grant No. 1111/17).
The work of TV is further supported by the European Research Council (ERC) under the EU Horizon 2020 Programme (ERC- CoG-2015 - Proposal n. 682676 LDMThExp), by the Israel Science Foundation-NSFC (grant No. 2522/17), by the German-Israeli Foundation (grant No. I-1283- 303.7/2014) and by a grant from The Ambrose Monell Foundation, given by the Institute for Advanced Study. This work was performed in part at the Aspen Center for Physics, which is supported by National Science Foundation grant PHY-1066293, and was partially supported by a grant from the Simons Foundation.

\appendix
\section{Full Boltzmann Equations for the 2-Component Model in the FIMP/Hot scenarios}

For completeness, we discuss the BEs for the 2-component model in the FIMP and thermalized sector scenarios, taking into account the 2-to-2 transfer terms. 

Following the notation used in \cite{Falkowski:2011xh}, 
the full Boltzmann equations for the FIMP case are 
\begin{align}\label{eq:BEfimpN}
 	\frac{d{Y}_N}{dz}
= -\frac{45}{2\pi^4 g_{*}} \gamma_N   z^3 K_1(z)
\Bigg( \frac{{Y}_N}{{Y}^{eq}_{N}} - \br_{l}  \Bigg),
 \end{align} 
\begin{align}\label{eq:BEfimpchi}
\frac{d{Y}_{\chi}}{dz}
=\frac{45}{2\pi^4 g_{*}}\gamma_N \br_{\chi } z^3 \Bigg[ \frac{{Y}_N}{{Y}^{eq}_{N}} K_1(z) 
+\frac{1}{2 \pi } \br_{l} U'(z)  \Bigg].
 \end{align} 
Where 
\beq
Y_N^{eq}(z)=\frac{g_N}{g_{*}}\frac{45}{4\pi^4} z^2 K_2(z), 
\eeq
$K_i(z)$ is the Bessel-K function of the $i^{\rm th}$ kind, and $\br_l$ is the branching fraction to the SM.
The first term in Eq.~(\ref{eq:BEfimpN}) describes decays of $N$'s to both sectors, while the second term represents inverse decays of SM states into $N$'s. Inverse decays from the hidden sector are neglected in the FIMP case ($f_{\chi,\phi}\to 0$). Eq.~(\ref{eq:BEfimpchi}) evolves the $\chi$ abundance. The source term proportional to the decay rate of $N$ is the same as in the approximate form we discussed earlier. 
The second term is due to transfer processes $\chi\phi \leftrightarrow l h$ (and all possible conjugations) between the 2 components, where once again, all effects of back-reaction from the hidden sector are taken to be feeble.
Following~\cite{Giudice:2003jh}, we use the on-shell subtracted transfer integral defined by 
\begin{align}
	U'(z) = U(z) - 3 \pi K_1(z).
\end{align}
Here, $U(z)$ is the full transfer integral, whereas the second term contains only the the pole contributions.
These must be subtracted in order to avoid double counting inverse decays of $l h \to N$. 

Explicitly, the transfer terms between the SM and DM sector take the form
\beq
U(z)\equiv\hat{\Gamma}_N \int_0^\infty d\hat{s}\sqrt{\hat{s}}\cdot K_1(z \sqrt{\hat{s}} )f_{s}(\hat{s},\hat{\Gamma}),
\eeq
where $\hat{\Gamma}_N\equiv\Gamma_N/M_N$, $s=(p_1+p_2)^2$ is the s-channel Mandelstam variable~\cite{Mandelstam:1958xc},  $\hat{s}=s/M_N^2$, and 
\beq
f_{s}(\hat{s},\hat{\Gamma})
\equiv   \frac{ \hat{s}^2 +2 \hat{s}}{(\hat{s}-1)^2+\hat{\Gamma}^2}.
\eeq

Working in the narrow width approximation, discussed in~\cite{Giudice:2003jh}, in which $\hat{\Gamma}_N\equiv\Gamma_N/M_N\to 0$, the pole in $f_s$ dominates the integral, and we find $U(z)\to 3\pi K_1(z)$, recovering the subtracted  on-shell component. 
Under this approximation, the BE reduces to the approximate form
\beq 
\frac{d{Y}_{\chi}}{dz}
=\frac{45}{2\pi^4 g_{*}}\gamma_N \br_{\chi } z^3 \frac{{Y}_N}{{Y}^{eq}_{N}} K_1(z) , 
\eeq 
and we may ignore the effects of any off-shell $2\leftrightarrow2$ processes.

In the thermalized sector case, we consider again the yield equation for the evolution of $N$, while the $\chi$ BE is replaced by the equation for the hidden sector total energy density $\rho_d$.
We then convert it, by taking the hidden sector to be thermalized, to an equation for $\xi(z)\equiv z/z'=T'/T$, evolving the temperature in the hidden sector measured with the SM clock. 
The Boltzmann equations for the thermalized case are
\bea \label{eq:hotN}
\frac{d{Y}_N}{dz}
&=& - \frac{45}{2\pi^4 g_{*}} \gamma_N   z^3 
\Bigg[ K_1(z) \Bigg(  \frac{{Y}_N}{{Y}^{eq}_{N}} - \br_{l} \Bigg) \nn \\
&-&  \br_\chi \xi K_1(z/\xi) \Bigg] , 
\eea
\bea \label{eq:hotdark}
\frac{d}{dz}\Big(\xi^4\Big)
&=&\gamma_N \br_\chi \Bigg[ \frac{4 }{3} \frac{g_{*}}{g'_{*,rh}}z^2 \Bigg( Y_N(z)-Y_N^{eq}(z/\xi) \Bigg)  \nn \\
& + &\frac{60}{  \pi^5 }\frac{1}{g'_{*,rh}}\br_{l} z^5\Bigg( \frac{I'(z)}{z}-\frac{I'(z/\xi)}{z/\xi}\Bigg) \Bigg].
\eea

The evolution of $N$, determined in this case by Eq.~(\ref{eq:hotN}), differs from Eq.~(\ref{eq:BEfimpN}) only by the third term, which describes inverse decays of hidden sector states into $N$'s with their own temperature $T'$.  
Eq.~(\ref{eq:hotdark}) evolves the dark sector temperature, with the source term proportional to the decay rate of $N$, as in the approximate form we discussed earlier. 
The second term decribes the backreaction through inverse decays of $\chi \phi \to N$, coming from the thermalized hidden sector, while the third and fourth terms are due to on-shell subtracted transfer processes $\chi\phi \leftrightarrow l h$  (and all possible conjugations) between the 2 components, where the effects from the hidden sector always come with the correct temperature dependence.
Here, the on-shell subtracted transfer integral is
\begin{align}\label{eq:Iprimedef}
	I'(z)=I(z)- 3\pi K_2(z),
	\end{align}
where $I(z)$ is the full transfer integral, and the $K_2(z)$ term represents the pole contributions. 
Explicitly, in this case, the transfer terms between the SM and DM sector take the form
where, similar to FIMP scenario, $I(z)$
\begin{align}\label{eq:Idef}
I(z)\equiv\hat{\Gamma}_N \int_0 ^{\infty} d\hat{s}\hat{s} \cdot K_2(\sqrt{\hat{s}}z) f_{s}(\hat{s},\hat{\Gamma}),
\end{align}
 using the same definitions for $\hat{s},f_s$ as before.
Taking again the limit $\hat{\Gamma}_N \to 0$, we find $I(z)\to 3\pi K_2(z)$, again recovering the on-shell  contribution. 
The BE then reduces to its approximate form
\begin{align}\label{eq:hotdarkapp}
 	\frac{d}{dz}\Big(\xi^4\Big)
=\gamma_N \br_\chi  \frac{2\pi^4 }{135} \frac{g_{*}}{g'_{*,rh}}z^2 \Bigg( Y_N(z)-Y_N^{eq}(z/\xi) \Bigg)\,.
 \end{align} 
The inverse decays from the hidden sector, as well as the energy transfer from the hidden sector to the SM, are suppressed by a factor of $\xi$ and can be neglected, unless the decay rates are large enough to thermalize the sectors. In that case, the full BEs are necessary to describe the thermalization of the two sectors correctly, as seen in Fig.~\ref{fig:BEsol}.

\section{ Derivation of the free streaming length}
We present the derivation of the free streaming length for the FIMP and hot DM scenarios, as well as the benchmark calculation often used for thermal relic DM.
Beginning with the definition of the free streaming length 
\begin{align}
	\mathcal{\lambda}_{\rm FS}	=\int_{t_{\rm rh}}^{t_{\rm eq}}\frac{\langle v\rangle}{a}dt
	=\int_{a_{\rm rh}}^{a_{\rm eq}}\frac{\langle v\rangle}{a^{2}H}da,
\end{align}
where $a(t_{\rm rh})$ is the scale factor at the end of leptogenesis, and $a_{\rm eq}=\Omega_{R}/\Omega_{M}$ is the scale factor at matter/radiation equilibrium.
This integral can be decomposed into two pieces
\begin{align}\label{eq:fsder}
	\mathcal{\lambda}_{\rm FS}	
	&= \int_{a_{\rm rh}}^{a_{\rm nr}}\frac{\langle v\rangle}{a^{2}H}da
	+ \int_{a_{\rm nr}}^{a_{\rm eq}}\frac{\langle v\rangle}{a^{2}H}da \nn \\
	&=\frac{1}{\sqrt{\Omega_{R}}H_{0}}\int_{a_{\rm rh}}^{a_{\rm nr}}da
	+\frac{a_{\rm nr}}{\sqrt{\Omega_{R}}H_{0}}\int_{a_{\rm nr}}^{a_{\rm eq}}\frac{da}{a\sqrt{1+a/a_{\rm eq}}},
\end{align}
describing the relativistic and non relativistic epochs of DM streaming. In Eq.~\ref{eq:fsder}, The effects of changes in $g_*$ on the Hubble rate, which are due to decoupling of relativistic degrees of freedom, can be neglected if the DM particle decouples while relativistic and streams for a sufficiently long period. Utilizing the Freidmann equation
\begin{align}
	a^{2}H=H_{0}\sqrt{\Omega_{R}}\sqrt{1+\frac{a}{a_{\rm eq}}}\,,	
\end{align}
the integral can be performed explicitly to give
\begin{align}
	\mathcal{\lambda}_{\rm FS}	&=
	 \frac{a_{\rm nr}}{\sqrt{\Omega_{R}}H_{0}}
	 \nn \\
	 &\times \left(1-2{\rm Arcsinh}(1) - \frac{a_{\rm rh}}{a_{\rm nr}}
	+2{\rm {\rm {Arcsinh}}}\sqrt{\frac{a_{\rm eq}}{a_{\rm nr}}}\right) 
	,
\end{align}
and in the limit $\frac{a_{\rm rh}}{a_{\rm nr}},\frac{a_{\rm nr}}{a_{\rm eq}}\ll1$ , the result is simplified to the familiar form
\begin{align}
	\mathcal{\lambda}_{\rm FS}
	& \simeq \frac{a_{\rm nr}}{\sqrt{\Omega_{R}}H_{0}}\left(0.624
	+{\rm ln}\left[\frac{a_{\rm eq}}{a_{\rm nr}}\right]\right).
\end{align}

Now, we are left with determining the value of $a_{\rm nr}$ for our various thermal DM production mechanisms.
We employ the fact that momentum redshifts as the scale factor to relate different times
\begin{align}
	 a_{\rm nr} = \frac{ \langle p \rangle_{\rm rh}}{\langle p \rangle_{\rm nr}} a_{\rm rh}, 
\end{align}
where different subscripts relate to thermally averaging at different temperatures,
and using the definition of the thermally averaged momentum
\begin{align}
	\langle p \rangle_T = 
	\frac{\int d^3 {\bf p} |{\bf p}| f({\bf p},T)}
	{\int d^3 {\bf p} f({\bf p},T)},
\end{align}
where $f({\bf p},T)$ is the phase space distribution at the relevant scale. 
We consider first the FIMP scenario. Assuming the sterile $N$'s decay while at rest, all the dark matter abundance is produced with $\langle p \rangle_{\rm rh} = M_N/2$. We relate $a_{\rm rh}$ to its value today by entropy dilution $a_{\rm rh} = a_0 \left( g_{*s,0} / g_{*s,\rm rh} \right)^{1/3} T_0/ T_{\rm rh}$. 
The average non-relativistic momentum can be defined by $\langle p \rangle_{\rm nr} =  m_\chi$,
and using the instantaneous decay approximation $\Gamma_N = H(T_{\rm rh}) \to \gamma_N=T_{\rm rh}^2/M_N^2$, these relations give
\begin{align}
	 a_{\rm nr}^{\rm FIMP} 
	 &= \frac{ M_N}{2 m_\chi} a_0 \left( \frac{g_{*s,0}} {g_{*s,\rm rh}} \right)^{1/3} \frac{T_0}{T_{\rm rh}} \nn \\
	 &= \frac{ T_0}{2 m_\chi}  \left( \frac{g_{*s,0}} {g_{*s,\rm rh}} \right)^{1/3} \gamma_N^{-1/2}.
\end{align}
Next, we consider the hot DM scenario.
In this case, there are several stages to consider. First, $\chi$ is produced and thermalizes quickly to $T'_{\rm rh}$. The hidden sector evolves as any degrees of freedom (different than $\chi$) within it decouple, thus changing its entropy. As the temperature decreases, the interactions maintaining $\chi$ in equilibrium cease, and $\chi$ freezes out of the hidden plasma while relativistic at $T'_{\rm fo}$. The final phase occurs when $\chi$ becomes nonrelativistic at $T'_{\rm nr}$. We can use the same method to compute $a_{\rm nr}$ by simply replacing $a_{\rm rh}$ with $a_{\rm fo}$
\begin{align}
	 a_{\rm nr} = \frac{ \langle p \rangle_{\rm fo}}{\langle p \rangle_{\rm nr}} a_{\rm fo}, 
\end{align}
and evolving forward in temperatures using entropy conservation in the hidden sector.
The phase space distribution at freeze-out is simply a FD distribution for a relativistic particle with $T' = \xi T$, so we have $\langle p \rangle_{\rm fo} = 3.15 \xi_{\rm fo} T_{\rm fo} $, and we may use 
$a_{\rm fo} = a_0 \left( g_{*s,0} / g_{*s,\rm fo} \right)^{1/3} T_0/ T_{\rm fo}$. 
By entropy conservation in the hidden and visible sectors, we find 
$\xi_{\rm fo} = \left( \frac{g'_{*s,\rm rh}   }{g'_{*s,\rm fo} } \frac{g_{*s,\rm fo}}  { g_{*s,\rm rh}} \right)^{1/3} \xi_{\rm rh}$. Again, by requiring $\langle p \rangle_{\rm nr} =  m_\chi$, we obtain
\begin{align}
	 a_{\rm nr}^{\rm Hot} 
	 &= \frac{ 3.15 T_0}{ m_\chi} \left( \frac{g_{*s,0}} {g_{*s,\rm rh}}
	 \frac{g'_{*s,\rm rh}}{g'_{*s,\rm fo}} \right)^{1/3} \xi_{\rm rh}.
\end{align}
As a sanity check, we also calculate the free streaming length for a thermal relic DM candidate. In this scenario, the DM candidate froze-out of equilibrium with the SM while relativistic. Essentially, this is the hot DM scenario with $\xi_{\rm fo}= \frac{g'_{*s,\rm rh}}{g'_{*s,\rm fo}}=1$, therefore 
\begin{align}
	 a_{\rm nr}^{\rm TR} 
	 &= \frac{3.15 T_0}{ m_\chi}  \left( \frac{g_{*s,0}} {g_{*s,\rm rh}} \right)^{1/3} 
	 ,
\end{align}
where $T_0=2.35\times10^{-4}~{\rm eV}$. We note that in this case $g_{*s,\rm rh}$ is a free parameter, since any dark degrees of freedom were assumed in equilibrium with the SM prior to decoupling.
One can express $g_{*s,\rm rh}$ in terms of the relic abundance
\begin{align}
	\Omega_{\rm DM}h^2
	& \simeq 
	\frac{Y_\chi^\infty s_0 m_\chi h^2}{\rho_c } 
	=0.12 \times \frac{135 \zeta(3)}{8 \pi^4}\frac{g_\chi}{g_{*s,\rm rh}}\frac{ h^2}{0.12 }
	\frac{m_\chi s_0} {\rho_c} \nn \\
	&=0.12 \times  \frac{g_\chi}{g_{*s,\rm rh}}\left(\frac{m_\chi}{2.2 \rm eV}\right)
\end{align}
arriving at
\begin{align}
	 a_{\rm nr}^{\rm TR} 
	 &= \frac{3.15 T_0}{ 2.2 \rm eV}  
	 \left( \frac{g_{*s,0}} {g_\chi} \right)^{1/3} 
	 \left( \frac{\Omega_{\rm DM}h^2}{0.12} \right)^{1/3}  
	 \left(\frac{2.2 \rm eV}{m_\chi}\right) ^{4/3},
\end{align}
giving, up to log corrections, the standard expression for the thermal relic free streaming length
\begin{align}
	\mathcal{\lambda}_{\rm FS}
	& \simeq 0.051~{\rm Mpc}\times
	\left( \frac{4.65~\rm keV}{m_\chi }\right)^{4/3}
	\left( \frac{g_{*s,0}} {3.91}  
	\frac{4}{g_\chi}
	 \frac{\Omega_{\rm DM}h^2}{0.12} \right)^{1/3}  .
\end{align}

\vspace{-6mm}

\bibliographystyle{apsrev4-1} 

\bibliography{LightDMbib.bib}

\end{document}